\documentstyle[PASJadd]{PASJ95}
\draft
\begin{document}

\title{Implication of Dark Matter in Dwarf Spheroidal
Galaxies}
\author{Hiroyuki {\sc Hirashita},$^1$\thanks{%
Research Fellow of the Japan Society for the Promotion of Science.}
\ Hideyuki {\sc Kamaya},$^{2*}$
and Tsutomu T. {\sc Takeuchi}$^{1*}$
\\[12pt]
$^1$ {\it Department of Astronomy, Faculty of Science, Kyoto University,
Sakyo-ku, Kyoto 606-8502}\\
{\it E-mail(HH): hirasita@kusastro.kyoto-u.ac.jp}\\  
$^2$ {\it Department of Physics, Faculty of Science, Kyoto
University, Sakyo-ku, Kyoto 606-8502}}

\abst{
We examine the correlation between physical quantities
to explore the existence of dark matter (DM) in the Local Group
dwarf spheroidal galaxies (dSphs). In order to clarify whether
DM exists in the dSphs we compare two extreme models of their
internal kinematics:
[1] a tidal model (the mass of a dSph is estimated with the luminous
mass, since the observed large velocity dispersions of the dSphs
are attributed to the tidal force by the Galaxy),
and
[2] a DM model (the mass of a dSph is estimated with the virial
mass, since the velocity dispersion is considered to reflect the
DM potential).
In both models, we find that the relation
between the surface brightness of the dSphs and the tidal force by
the Galaxy is consistently interpreted. This makes a critical
remark about the previous studies
that concluded that the tidal force is effective for the dSphs
based only on model [1]. We also check the correlation between the
Galactocentric distance and the
tidal force. Consequently, both models are also supported
in this test. Thus, we are unable to judge which of the two is more
promising for the dSphs in correlation investigations.
The physical process of tidal disruption is also considered.
Since the timescale for the tidally perturbed stars to
escape is much shorter than the orbital periods
of the dSphs, it is difficult for the tidally perturbed dSphs to
exist as an assembly. Thus, we suggest that the
gravitational field of DM is necessary to bind the dSphs with a
large velocity dispersion.
}

\kword{Dark matter --- Galaxies: elliptical ---
Galaxies: evolution}

\maketitle
\thispagestyle{headings}

\section{Introduction}

Recent observations have been revealing the properties of the Local
Group dwarf spheroidal galaxies (dSphs). The dSphs have luminosities
of order
$10^5$--$10^7\,\LO$, and are characterized by their
low surface brightnesses (Gallagher, Wyse 1994 for review).
The observations of such faint objects are
important for several reasons. For example, in a galaxy-formation
theory based on the cold dark-matter (CDM) model,
low-mass galaxies are considered to be the first bound luminous
objects (e.g., Blumenthal et al.\ 1984). Thus, the dSphs are
expected to have
a large amount of the CDM (see also Flin 1999 for an observational
approach).

To date, the stellar velocity dispersions of dSphs have been
extensively measured
(e.g., Mateo et al.\ 1993 and references therein),
which, in general, indicate a too large mass to be accounted for
by the visible stars in the dSphs. In other words,
dSphs generally have high mass-to-light ratios ($M_{\rm vir}/L_V$,
where $M_{\rm vir}$ is the virial mass calculated from observed
size and stellar velocity dispersion and $ L_V$ is the total
luminosity at the $V$ band;
the 6th column in table 1). This fact
can imply the presence of dark
matter (DM) in these systems (e.g., Mateo et al.\ 1993).
The existence of
DM is also supported by the large spatial distribution of stars in
their outer regions (Faber, Lin 1983; Irwin, Hatzidimitriou 1995).


However, the above arguments may be challenged if we take into account
the tidal force exerted by the Galaxy.
If a dwarf galaxy orbiting the Galaxy is
significantly perturbed by the Galactic tidal force, the observed
velocity dispersion of the dwarf
galaxy can be larger than the gravitational equilibrium dispersion
(Kuhn, Miller 1989; Kroupa 1997). Moreover, the
results in Kuhn (1993) imply that the lifetimes of internally unbound
satellites might be longer than one orbital period (but see Johnston
1998). The timescale
for such an unbound object to survive is estimated later in this
paper (subsubsection 3.2.1). This tidal picture of the dSphs also
suggests that the large velocity dispersions do not
necessarily show the existence of DM (but see
Piatek, Pryor 1995; Oh et al.\ 1995).
We will consider whether the observed large mass-to-light ratios of
dSphs are really due to the tidal effect.

The plan of this paper is as follows.
First of all, in the next section, we present correlations among the
observed quantities.  We summarize our results in section 3. In the
same section,
some discussions are also made, and we comment about the
environmental effects on the dSphs.

\section{Correlation among Observed Quantities}

\subsection{Surface Brightness and Galactocentric Distance}

In this section, we consider what determines the physical condition
of the dSphs. We re-examine the correlation which
Bellazzini et al.\ (1996; hereafter B96) have presented.

First, a dimensionless form of the Galactic tidal force is defined.
The Galactic potential is modeled by a spherical dark halo with a
flat rotation curve of amplitude $V_{\rm rot}$
(Honma, Sofue 1997 for the latest results). We assume that the
dark halo extends up to $R_{\rm GC}=100$ kpc.
With the above flat-rotation model of the Galactic potential, the
mass of the Galaxy within the
Galactocentric distance $R_{\rm GC}$ is expressed as
\begin{eqnarray}
M_{\rm G}(R_{\rm GC})=\left\{
\begin{array}{ll}
{\displaystyle \frac{V_{\rm rot}^2R_{\rm GC}}{G}}=1.1\times
10^{12}\MO\left({\displaystyle \frac{R_{\rm GC}}{100\,{\rm kpc}}}
\right) &
\mbox{for} \ R_{\rm GC}\leq 100\,\mbox{kpc}, \\
1.1\times10^{12}\MO & \mbox{for} \
R_{\rm GC}\geq 100\,\mbox{kpc},
\end{array}
\right.
\end{eqnarray}
where we assumed $V_{\rm rot}=220\,\mbox{km s}^{-1}$
(e.g., Burkert 1997).

We consider a spherical satellite galaxy orbiting the Galaxy.
The tidal force which a star (whose mass is $m_*$) at the core
radius ($r_{\rm c}$; the radius where the surface brightness falls
to half its central value; Binney, Tremaine 1987, p.\ 25) of the
satellite galaxy experiences is approximately
\begin{eqnarray}
F_{\rm T}\equiv\frac{Gm_{*}M_{\rm G}r_{\rm c}}{R_{\rm GC}^3}.
\label{tidal}
\end{eqnarray}
The star also experiences a gravitational binding force exerted
by the satellite galaxy, which is described as
\begin{eqnarray}
F_{\rm B}\equiv\frac{Gm_*M_{\rm s}}{r_{\rm c}^2},
\end{eqnarray}
where $M_{\rm s}$ is the total mass of the satellite galaxy. We
then define a dimensionless tidal force identical to the
definition by B96,
\begin{eqnarray}
F_{\rm T,B}\equiv\frac{F_{\rm T}}{F_{\rm B}}=
\left(\frac{M_{\rm G}}{M_{\rm s}}\right)\left(
\frac{r_{\rm c}}{R_{\rm GC}}\right)^3.\label{ftb}
\end{eqnarray}
We estimate $M_{\rm s}$ in the following two different ways:

\noindent
[1] Tidal model: The large velocity dispersion of dSphs
is attributed not to deep DM potentials, but to significant
perturbation by the Galactic tides.

\noindent
[2] DM model: A significant amount of DM exists in a dSph, and
virial equilibrium in the DM potential is realized.

\noindent
For the tidal model, the mass of every dSph is estimated in the
following way:
\begin{eqnarray}
M_{\rm s}=M_{\rm L},
\end{eqnarray}
where $M_{\rm L}$ is the luminous mass calculated directly from the
total luminosity at the $V$ band by assuming $ M_{\rm L}/L_V=1$
(in solar units). The
difference in the assumed value of $ M_{\rm L}/L_V$ does not affect the
analyses in this paper as long as it is constant. For the DM model,
\begin{eqnarray}
M_{\rm s}=M_{\rm vir},
\end{eqnarray}
where $M_{\rm vir}$ is the virial mass derived from the observed
central velocity dispersion by assuming virial equilibria
(Pryor, Kormendy 1990; 4th column of table 1). We assume in the
DM model (i.e., when
estimating the virial mass) that mass
traces light (e.g., Pryor 1992). Though we cannot
judge by the data available at present whether this assumption is
true for dSphs, we know that mass does not follow light in more
luminous galaxies (e.g., Binney, Tremaine 1987, chap.\ 2). However,
the assumption has been made by many previous investigations, so
we make it in order to make comparisons with them.

Figure 1 shows the correlation between $\mu_V(0)$ (central surface
brightness in $V$ band) and
$\log F_{\rm T,B}$ for the tidal model.
We calculate the linear correlation coefficient,
$r$, by
\begin{eqnarray}
r=\frac{\sum_i(x_i-\bar{x})(y_i-\bar{y})}
{\sqrt{\sum_i(x_i-\bar{x})^2}\sqrt{\sum_i(y_i-\bar{y})^2}},
\end{eqnarray}
where $(\bar{x},\,\bar{y})$ is the average of all data points
$(x_i,\, y_i)$ and $\sum_i$ means the summation of all the data.
The correlation coefficient is
$r=0.91$ for the data points of the tidal model, which indicates
a strong correlation. {(If $r>0.71$ for 8 data, the two quantities
are correlated with a confidence level of 95\%.)} Thus, B96
insisted that the Galactic tidal effect had settled the present
ranking in the dSph central brightness. We have indeed confirmed
this result.


Here, the DM model is tested by the same method as
B96, except for mass estimations. We derive the mass by using the
virial theorem, $M_{\rm s}=M_{\rm vir}$. Figure 2 shows the
relation between $\mu_V(0)$ and
$F_{\rm T,B}$ for the DM model. In this model, we find that the two
quantities are less correlated ($r=0.48$) than in the tidal model.
A lower correlation is naturally expected for the DM model,
since it is the DM in a dSph, not the tidal force of the Galaxy,
that determines the dynamical condition of the dSphs.


We note that B96 only tested the tidal model.
Since the DM model as well as the tidal model can be consistently
interpreted by the same method as B96, we cannot conclude that
the tidal force is effective in determining the physical conditions
for the dSphs. Indeed, B96 commented in their subsection 3.3 that
their analyses do not necessarily show that that tidal force
is important for the dSphs.

Thus, there is nothing to choose between the two models in the
$\mu_V(0)$--$F_{\rm T,B}$ relations. We, therefore, need further
considerations to determine which of the two models is applicable
for dSphs. The further examination is discussed in the
next subsection.

\subsection{Tidal Force and Galactocentric Distance}

In this subsection we present a correlation between the
dimensionless tidal force ($\log F_{\rm T,B}$) and the Galactocentric
distance ($\log R_{\rm GC}$) for each model. This correlation is
shown in figures 3 and 4 for the tidal model and the DM model,
respectively.


\subsubsection{The $log\, F_{T,B}$--$log\, R_{GC}$
relation: results}

First, the $\log F_{\rm T,B}$--$\log R_{\rm GC}$
relation is plotted in figure 3 using the tidal model
($M_{\rm s}=M_{\rm L}$). In
this model, the correlation coefficient
is $r=-0.94$ and the linear regression is
\begin{eqnarray}
\log F_{\rm T,B}=-4.12\log R_{\rm GC}+6.55.
\end{eqnarray}
We obtain a strong correlation between the two quantities.
One interpretation of the regression is that the
tidal force is more effective for dSphs nearer to the
Galaxy. However, this interpretation is not
unique, as shown in subsubsection 2.2.2.

Next, the same relation is presented in figure 4 by using the DM model
($M_{\rm s}=M_{\rm vir}$). The two quantities are marginally
correlated ($r=-0.68$).
The linear regression for this model becomes
\begin{eqnarray}
\log F_{\rm T,B}=-1.50\log R_{\rm GC}-0.105.\label{reg2}
\end{eqnarray}

\subsubsection{Interpretation of the regressions}

In the tidal model, the correlation presented in the previous
subsubsection is stronger than that in the DM model. 
The regression and the stronger correlation in the tidal model
 between $F_{\rm T,B}$ and  $R_{\rm GC}$ (figure 3) may mean that
the tidal forces
are more effective for dSphs located nearer to the Galaxy
(subsubsection 2.2.1),
but there is a known correlation between the luminosity and
the Galactocentric distance, both of which are directly observable.
Since we determine the mass of a dSph by luminosity in the tidal
model, the strong correlation can reflect a correlation between
luminosity and the Galactocentric distance. Indeed,
B96 (their figure 1) showed
that $\mu_V(0)$ and $R_{\rm GC}$ are correlated.
Because $F_{\rm T,B}$, comprising a few observed quantities, is
not directly observable, the good correlation
between $F_{\rm T,B}$ and $R_{\rm GC}$ has only a secondary meaning,
showing one of the interpretations on the relation among
observed quantities.

According to Caldwell et al.\ (1998), the relations among the
surface brightness, luminosity, and metallicity of dSphs belonging
to the M81 Group show no difference from the relations of the
Local Group dSphs. This may imply that the physical properties of
dSphs are determined by their intrinsic conditions, not by their
environments. However, there is still a possibility that the
environments influence the dSphs so that the relations among
surface brightness, luminosity, and metallicity
are conserved.

In the DM model, we find that in equation (\ref{reg2})
\begin{eqnarray}
F_{\rm T,B}\propto R_{\rm GC}^{-1.50}.\label{modelpre}
\end{eqnarray}
The error ($\sigma_{\rm c}$) of the slope of regression
(\ref{reg2}) is estimated from the propagation of error
(e.g., Hoel 1971),
\begin{eqnarray}
\sigma_{\rm c}=\sqrt{\frac{n\sigma_F^2}{\Delta}},
\end{eqnarray}
where $n$ and $\sigma_F^2$ mean the number of data and
the dispersion of data $\log F_{\rm T,B}$, respectively.
Here, $\Delta$ is defined by
\begin{eqnarray}
\Delta\equiv n\sum_{i=1}^{n}R_i^2-\left(\sum_{i=1}^nR_i
\right)^2.
\end{eqnarray}
The data set $\{R_i\}_{i=1}^n$ is composed of $n$ (here, $n=8$)
data of $R_{\rm GC}$; $\{R_i\}_{i=1}^n$.

Using directly observable quantities, $F_{\rm T,B}$ is
expressed as
\begin{eqnarray}
F_{\rm T,B}\propto\frac{\theta^2}{v^2},
\end{eqnarray}
where we used the relations $M_{\rm G}\propto R_{\rm GC}$,
$M_{\rm s}\propto r_{\rm c}v^2$ (virial equilibrium), and
$\theta = r_{\rm c}/R_{\rm GC}$ ($v$ and $\theta$ mean the
internal velocity dispersion and the angular size of a dSph,
respectively). Assuming the error of $\theta$ and $v$ to be
10\% and 20\%, respectively (Mateo et al.\ 1993), $\sigma_F$ is
estimated as $\sigma_F\simeq 0.24$. This leads to
$\sigma_{\rm c}\simeq 0.40$. Thus, we obtain the regression
\begin{eqnarray}
F_{\rm T,B}\propto R_{\rm GC}^{-1.50\pm 0.40}.\label{flat}
\end{eqnarray}

If the flat rotation curve of the Galaxy is taken into account,
$M_{\rm G}\propto R_{\rm GC}$. Using equation (\ref{ftb}),
we obtain
\begin{eqnarray}
F_{\rm T,B}\propto\rho_{\rm s}^{-1}R_{\rm GC}^{-2},\label{density}
\label{theorypre}
\end{eqnarray}
where we define the mass density of satellite dwarf galaxies as
\begin{eqnarray}
\rho_{\rm s}=\frac{M_{\rm s}}{r_{\rm c}^3}.\label{rhos}
\end{eqnarray}
If the density of a dSph is determined by DM contained in
itself, $\rho_{\rm s}$ is independent of $R_{\rm GC}$.
Using relations (\ref{flat}) and (\ref{density}), we obtain
\begin{eqnarray}
\rho_{\rm s}\propto R_{\rm GC}^{-0.50\pm 0.40}.
\end{eqnarray}
This shows that it cannot be determined whether $\rho_{\rm s}$
depends on $R_{\rm GC}$ strongly or weakly.
We need a larger sample of galaxies to judge whether the density of
the dSphs is internally or environmentally determined.

Burkert (1997) also examined the relation between the mass
density and the Galactocentric distance with a smaller sample
than ours. This argument supports both the tidally perturbed
picture and the DM-dominated picture of the dSphs.

In summary, we have no
evidence which model gives a better kinematical picture of the
dSphs. This implies that a statistical study of the physical
quantities of the dSphs may be unable to prove the existence of
DM in the dSphs. Thus, a straightforward examination of the
physical process of tidal disruption is necessary.
The subject of tidal heating has been investigated in many
researches (e.g., Weinberg 1994; Kundi\'{c}, Ostriker 1995;
Gnedin, Ostriker 1997).

\section{Summary and Discussions}

\subsection{Summary}

We re-examined the correlations between the physical quantities
of the Local Group dSphs. We estimated the mass of the dSphs by
the following two ways:

\noindent
[1] tidal model (the mass of a dSph is estimated with the luminous
mass), and

\noindent
[2] DM model (the mass of a dSph is estimated with the virial mass).

In the tidal model, there is strong correlation between the surface
brightnesses of the dSphs and the tidal force by the Galaxy,
as shown by B96. In the DM
model, as expected, a correlation is barely found between the
two quantities. This
is consistent for the DM model, since the internal kinematic
conditions of dSphs are determined only by their internal DM.
We also examined the correlation between the Galactocentric distance
and the dimensionless tidal force defined in equation (\ref{ftb}).
Here, we obtained a strong correlation in the tidal model, while we
found a weak correlation in the DM model. For the two models,
the strengths of the derived correlations support both models.
In other words,
we cannot determine which of the two is superior by examining the
correlations mentioned above.
We note that B96 does not necessarily assert that the
tidal force is the dominant factor for determining the 
kinematical conditions in the dSphs (their subsection 3.3).
However, many papers cite
B96 to show that the tidal forces determine the structure
of the dSphs based only on their subsections 2.3--2.4 and figure 3.
We stress again that we can never judge the
importance of the tidal effect from figure 3 of B96.

Our approach is an alternative to the discussion by
Burkert (1997), who compared the DM-dominated picture
and tidally perturbed picture of the Local Group dSphs using
a smaller sample than ours.
We have demonstrated that the superiority of the model can
never be judged from statistical analyses of the presently
available data. Thus, we should resort to methods other than
statistical methods.

\subsection{Discussions}

\subsubsection{Tidal origin of the large velocity dispersions?}

If circular orbits are assumed for the dSphs, the tidal forces
from the Galaxy are
constant in time. Both figures 3 and 4 show
that $F_{\rm T,B}\ltsim 0.1$ for all dSphs, even when the tidal
model is adopted. Hence, the effect of tidal
heating proves to be too small to account for the large velocity
dispersions in the dSphs if the orbits are circular.
The same view is found in figure 1 of Pryor (1996).

Tidal heating may be effective for the elliptical orbit
with a sufficiently small perigalactic distance. However,
Piatek and Pryor (1995) showed by numerical simulations that the
velocity dispersions of dSphs are little affected by tidal forces,
even when the tidal
forces effectively disrupt the outer regions of the dSphs.
Moreover, the escape timescale, $t_{\rm esc}$, defined by the time
needed for the tidally
heated star to escape to $l=1$ kpc from a dSph, is estimated as
\begin{eqnarray}
t_{\rm esc}=10^8\, \left(\frac{l}{1\,\mbox{kpc}}\right)
\left(\frac{v_{\rm esc}}{10\,\mbox{km s}^{-1}}\right)^{-1}\,
\mbox{yr},
\end{eqnarray}
where $v_{\rm esc}$ is the escape velocity of the dSph.
This estimate is consistent with that of Johnston (1998).
The estimated escape velocity is an order of magnitude smaller than
the orbital timescale of dSphs ($\sim 1$ Gyr). Thus, the
significantly tidal-perturbed stars
in dSphs go their separate ways and cannot be observed as an
assembly. Therefore, we suggest that tidal disruption seems to be
unlikely to explain the large velocity dispersions of the dSphs.

There have been some simulations which indicated that the dSph
systems are tidally disrupted in the age of the Universe by the
collective effect of successive passages of the pericenter
(Kuhn, Miller 1989; Oh et al.\ 1995; Kroupa 1997). We note
that in the early epoch of the Universe, when the scale factor of
the Universe is smaller than that of today and the expansion
of the Universe is not negligible, the distance between a dSph
and the Galaxy is smaller, which leads to a stronger tidal
force. Thus, further simulation is needed to take into
account the expansion effect (see Mishra 1985 for treatment of
the cosmic expansion). If such an effect is included, the dSphs
may be quickly disrupted in the galaxy-formation epoch without the
presence of the DM to bind themselves.

For further discussions and reviews concerning the tidal heating of
dSphs, see Pryor (1996).

\subsubsection{Other environmental effects?}

The ram pressure of the gaseous Galactic halo may have an
influence on the dSphs. Here, we briefly estimate the ram pressure
force of the halo gas acting on
the orbiting dSphs. 

Songaila (1981) estimated the
mass density of the gaseous halo to be
$\rho_{\rm halo}\sim 10^{-27}\,\mbox{g cm}^{-3}$. Using this value,
the ram pressure force by the gaseous halo is
estimated as
\begin{eqnarray}
\rho_{\rm halo}V_{\rm rot}^2(\pi r_{\rm c}^2)\sim 10^{30}\left(
\frac{\rho_{\rm halo}}{10^{-27}\,\mbox{g cm}^{-3}}\right)
\left(\frac{V_{\rm rot}}{220\,\mbox{km s}^{-1}}\right)^2\left(
\frac{r_{\rm c}}{300\,\mbox{pc}}\right)^2\,\mbox{dyn}.
\end{eqnarray}
Since the tidal force by the Galaxy at the perigalactic distance
$R_{\rm p}$ is estimated as (equation \ref{tidal})
\begin{eqnarray}
\frac{GM_{\rm G}M_{\rm s}r_{\rm c}}{R_{\rm p}^3}=
\frac{V_{\rm rot}^2M_{\rm s}r_{\rm c}}{R_{\rm p}^2}\sim 10^{30}
\left(\frac{M_{\rm s}}{10^7\MO}\right)\left(
\frac{r_{\rm c}}{300\,\mbox{pc}}\right)\left(
\frac{R_{\rm p}}{30\,\mbox{kpc}}\right)^{-2}\,\mbox{dyn} ,
\end{eqnarray}
the ram pressure is comparable to the tidal force. 
This means that even in the situation where the tidal force
is effective, the ram pressure is also effective. Therefore, we
should not conclude that only the tidal force has established the
present surface brightnesses of the dSphs.
It may be possible that the density gradient
of the gaseous Galactic halo may produce the
correlation which B96 presented.
Einasto et al.\ (1974) pointed out the correlation between
the morphological types of companion galaxies and their distances
from the parent galaxies. They also suggested that the density of
coronal gas in the halo 
$\rho_{\rm c}$ is proportional to $R_{\rm GC}^{-2}$.

Actually, we need a more precise treatment of the ram pressure,
which takes into account various factors for dSphs; the fraction
of the gas mass to the total mass, the internal distribution of
the gas, the form of the gas (dense or diffuse state), etc. The
process of ram-pressure stripping is extensively investigated by
numerical simulations by Portnoy et al.\ (1993). The application
of this process to
the Galactic halo should be addressed by future work.

A correlation may be produced in the formation epoch of the Local
Group. For example, a Galactic wind in its initial
starburst phase may affect the star-formation histories of dSphs
(van den Bergh 1994; Hirashita et al.\ 1997).
Since different star-formation
histories mean different `supernova histories,' and supernovae may
have played a dominant role in determining
the structure parameters (for example, luminosity profiles:
Larson 1974; Saito 1979; see also Hirashita 1999), the
difference in star-formation histories may lead to the correlation
which B96 pointed out.

We note that in the DM context of the dSphs, the suggestion by
Hirashita et al.\ (1997) has another meaning. According to
Mac Low and Ferrara (1999), dwarf galaxies cannot necessarily
lose all of their gas by supernova heating
in the presence of a DM potential. Hence, the
ram pressure by the Galactic wind may be necessary for
gas depletion in the dSphs.

\subsubsection{DM in dSphs}

\noindent
[1] Although we concluded in subsubsection 3.2.1 that the large
mass-to-light ratio is not explained by only the tidal force, it
becomes possible if the
DM content is considered. Considering that tidally disrupted dSphs
survive only $\sim 10^8$ yr (subsubsection 3.2.1), the
DM content in dSphs is needed to bind a dSph as a system.
As implied in subsubsection 3.2.2,
ram-pressure stripping by the Galactic gaseous halo or supernovae
in the dSphs may be effective for determining the physical
parameters of the dSphs. Even in this context, the DM content is 
necessary to maintain the dSphs as a bound system.

\noindent
[2] In section 2, the linear regression (\ref{reg2}) and correlation
coefficient in figure 4 indicate that $\rho_{\rm s}$ may correlate
a little with the Galactocentric distance (subsubsection 2.2.2).
Since this implies that the density is
determined by the internal structure of the dSphs, this is
consistent with the DM-dominant picture of dSphs (Oh et al.\ 1995).

Here, we note that Hirashita et al.\ (1998) explained
the mass-luminosity relation of the dwarf spheroidal galaxies in the
context of galaxy formation in the DM potential.
Finally, we should stress again that the analyses by B96
necessarily showed the dominance of the tidal force in determining
the kinematical conditions of the Local Group dSphs.
Considering the above [1] and [2] as well as the result in
section 2,  the DM-dominated dSphs are no less probable than the tidally
perturbed dSphs.

\par
\vspace{1pc}\par
We first thank Prof.\ C.\ Pryor, the referee, for his careful
reading and making valuable comments, which improved this paper
very much. We are grateful to Profs.\ S.\ van den Bergh,
T.\ E.\ Armandroff, M.\ Bellazzini for useful comments
about environmental effects on satellite galaxies. Their comments 
at the IAU meeting were helpful to improve the discussions in the
paper. We also thank Profs.\ M.\ Sait\={o}, Drs.\ S.\ Mineshige,
and T.\ T.\ Ishii for helpful
comments. All of us acknowledge the Research
Fellowship of the Japan Society for the Promotion of Science for
Young Scientists. We made extensive use of the NASA's Astrophysics
Data System Abstract Service (ADS).

\section*{References}
\small
 
\re
Bellazzini M., Fusi Pecci F., Ferraro F.R.\ 1996, MNRAS
278, 947 (B96)
\re
Binney J., Tremaine S.\ 1987, Galactic Dynamics (Princeton University
Press, Princeton)
\re
Blumenthal G.R., Faber S.M., Primack J.R., Rees M.J.\
1984, Nature 311, 517 
\re
Burkert A.\ 1997, ApJ 474, L99
\re
Caldwell N., Armandroff T.E., Da Costa G.S., Seitzer P.\ 1998, AJ
115, 535
\re
Caldwell N., Armandroff T.E., Seitzer P., Da Costa G.S.\
1992, AJ 103, 840
\re
Einasto J., Saar E., Kaasik A.\ 1974, Nature 252, 111
\re
Faber S.M., Lin D.N.C. 1983, ApJ 266, L17
\re
Flin P.\ 1999, in The Stellar Content of Local
Group Galaxies, ed P. Whitelock, R. Cannon (Kluwer, Dordrecht)
in press
\re
Gallagher J.S. III, Wyse R. F. G.\ 1994, PASP 106, 1225
\re
Gnedin O.Y., Ostriker J.P.\ 1997, ApJ 474, 223
\re
Hirashita H.\ 1999, ApJ in press
\re
Hirashita H., Kamaya H., Mineshige S.\ 1997, MNRAS
290, L33
\re
Hirashita H., Takeuchi T.T., Tamura N.\ 1998, ApJ 504, L83
\re
Hoel P.G.\ 1971, Introduction to Mathematical
Statistics, 4th edition (Wiley, New York) ch7
\re
Honma M., Sofue Y.\ 1997, PASJ 49, 453
\re
Irwin M., Hatzidimitriou D.\ 1995, MNRAS 277, 1354
\re
Johnston K.V.\ 1998, ApJ 495, 297
\re
Kroupa P.\ 1997, New Astron.\ 2, 139
\re
Kuhn J.R.\ 1993, ApJ 409, L13
\re
Kuhn J.R., Miller R.H.\ 1989, ApJ 341, L41
\re
Kundi\'{c} T., Ostriker J. P.\ 1995, ApJ 438, 702
\re
Larson R.B.\ 1974, MNRAS 169, 229
\re
Lee M.G., Freedman W., Mateo M., Thompson I., Roth M.,
Ruiz M.-T.\ 1993, AJ 106, 1420 
\re
Mac Low M.-M., Ferrara A.\ 1999, ApJ 513, 142
\re
Mateo M., Olzewski E.W., Pryor C., Welch D.L., Fischer P.\
1993, AJ 105, 510
\re
Mishra R.\ 1985, MNRAS 212, 163
\re
Oh K.S., Lin D.N.C., Aarseth S.J.\ 1995, ApJ 442, 142
\re
Piatek S., Pryor C.\ 1995, AJ 109, 1071
\re
Portnoy D., Pistinner S., Shaviv G.\ 1993, ApJS 86, 95
\re
Pryor C.\ 1992, in Morphological and Physical Classification of
Galaxies, ed G.\ Longo, M.\ Capaccioli, G.\ Busarello
(Kluwer, Dordrecht) p163
\re
Pryor C.\ 1996, in Formation of the Galactic Halo, ed H.\ Morrison, 
A.\ Sarajedini, ASP Conf.\ Ser.\ 92, p424
\re
Pryor C., Kormendy J.\ 1990, AJ 100, 127
\re
Saito M.\ 1979, PASJ 31, 193
\re
Songaila A.\ 1981, ApJ 248, 945
\re
van den Bergh S.\ 1994, ApJ 428, 617
\re
Vogt S.S., Mateo M., Olszewski E.W., Keane M.J.\ 1995,
AJ 109, 151
\re
Weinberg, M. D.\ 1994, AJ  108, 1414
\re
Zaritsky D., Olszewski E.W., Schommer R.A., Peterson R.C.,
Aaronson M.\ 1989, ApJ 345, 759 

\label{last}

\clearpage

\begin{table}[t]
\small
\begin{center}
Table~1.\hspace{4pt}Local Group dwarf spheroidal galaxies.$^*$
\end{center}
\vspace{6pt}
\begin{tabular*}{\columnwidth}
{@{\hspace{\tabcolsep}
\extracolsep{\fill}}p{5pc}lccccccc}
\hline\hline\\[-6pt]
Name & $R_{\rm GC}$ & $L_V$ & $r_{\rm c}$ & $M_{\rm vir}$ &
$M_{\rm vir}/L_V$ & $\mu_V(0)$ & Ref$^\dagger$
 \\
           & (kpc) & ($10^5\, \LO$) & (pc) &
($10^7\, \MO$) &
(Solar units) & (mag arcsec$^{-2}$) & \\ \hline
Ursa Minor \dotfill & 66 & 3.0 & 290 & 3.9   & 130 & 25.1 & 1 \\
Draco      \dotfill & 76 &   2.5 & 190 & 5.2   & 210 & 25.2 & 1 \\
Sculptor   \dotfill & 78 &   16  & 200 & 1.4   & 8.8 & 24.1 & 1 \\
Carina     \dotfill & 89 &   2.9 & 210 & 1.1   & 38  & 25.2 & 1 \\
Sextans    \dotfill & 91 &   8.3 & 380 & 2.6   & 31  & 25.5 & 1 \\
Fornax     \dotfill & 133 &  250 & 640 & 12    & 4.8 & 23.2 & 1 \\
Leo II     \dotfill & 219 &  9.9 & 220 & 1.1   & 11  & 24.0 & 2 \\
Leo I      \dotfill & 270 &  40.9 & 260 & 0.47 & 1.2 & 22.3 & 3, 4, 5 \\
[4pt] \hline
\end{tabular*}
\vspace{6pt}\par\noindent
$*$  See text for the definitions of quantities.\\
$\dagger$ (1) Mateo et al.\ 1993; (2) Vogt et al.\ 1995;
(3) Lee et al.\ 1993 (4) Caldwell et
al. 1992; (5) Zaritsky et al.\ 1989.
\end{table}

\clearpage

\bigskip

\begin{fv}{1}
{7cm}
{Relation between $\mu_V(0)$ (central surface brightness in the $V$
band) and $F_{\rm T,B}$ (dimensionless tidal force) for the tidal
model for each dSph in the Local Group. In this model, the mass of
each dSph is estimated by the
luminous mass. The correlation coefficient is 0.91.}
\end{fv}
\begin{fv}{2}
{7cm}
{Same as figure 1, but for the dark-matter model. The mass of
each dSph is estimated by the virial mass. The correlation
coefficient is 0.48.}
\end{fv}
\begin{fv}{3}
{7cm}
{Correlation between $F_{\rm T,B}$ (dimensionless tidal force) and
$R_{\rm GC}$ (Galactocentric distance) for the tidal model.
The correlation coefficient is $-0.94$. The line represents the
linear regression.}
\end{fv}
\begin{fv}{4}
{7cm}
{Same as figure 3, but for the dark-matter model. The
correlation coefficient is $-0.68$.}
\end{fv}
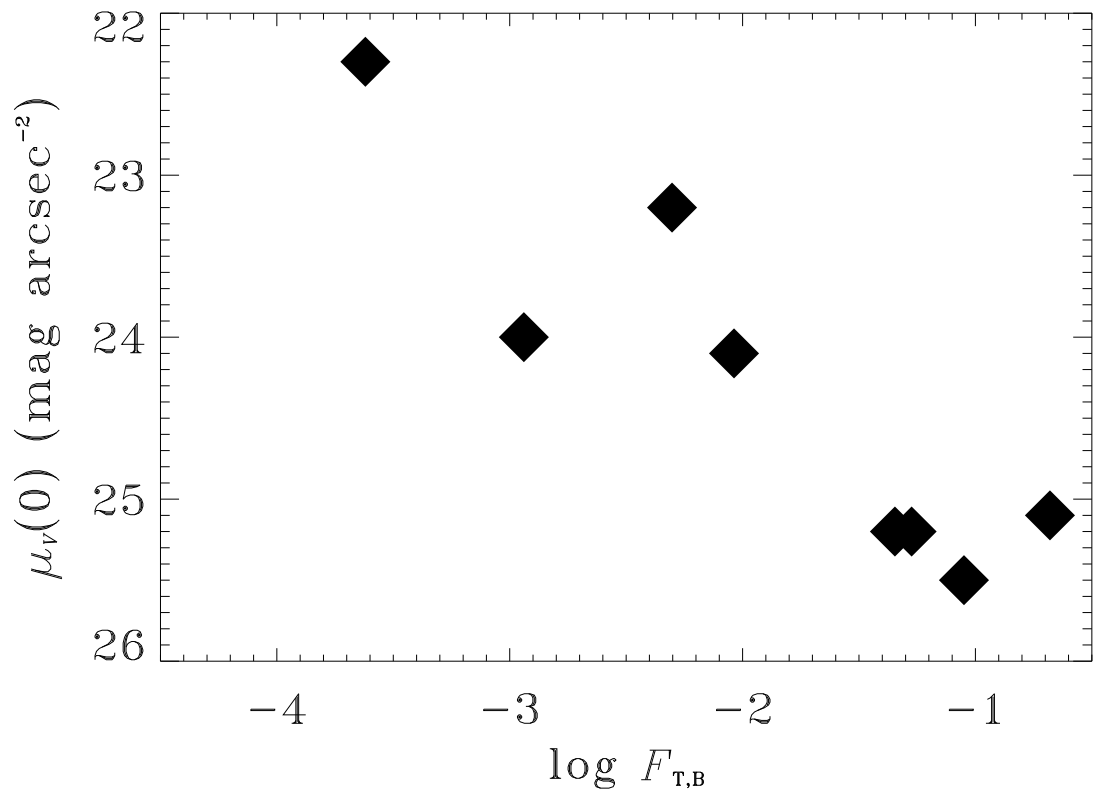
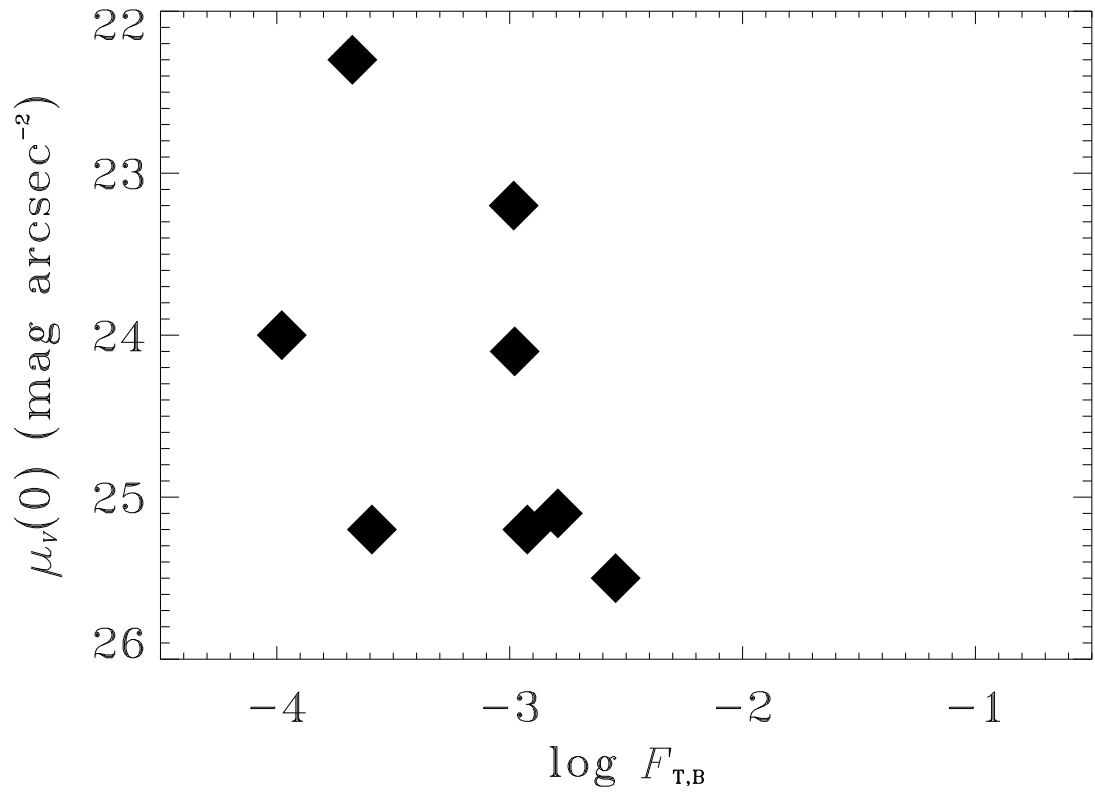
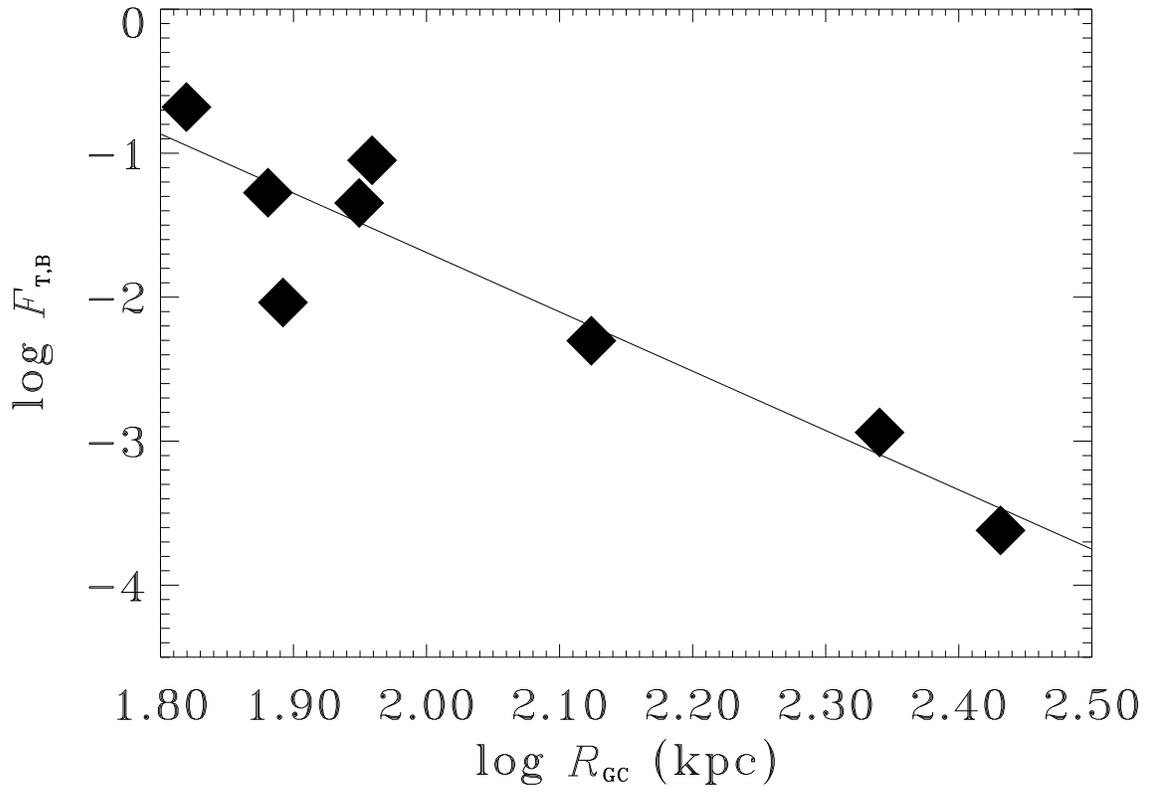
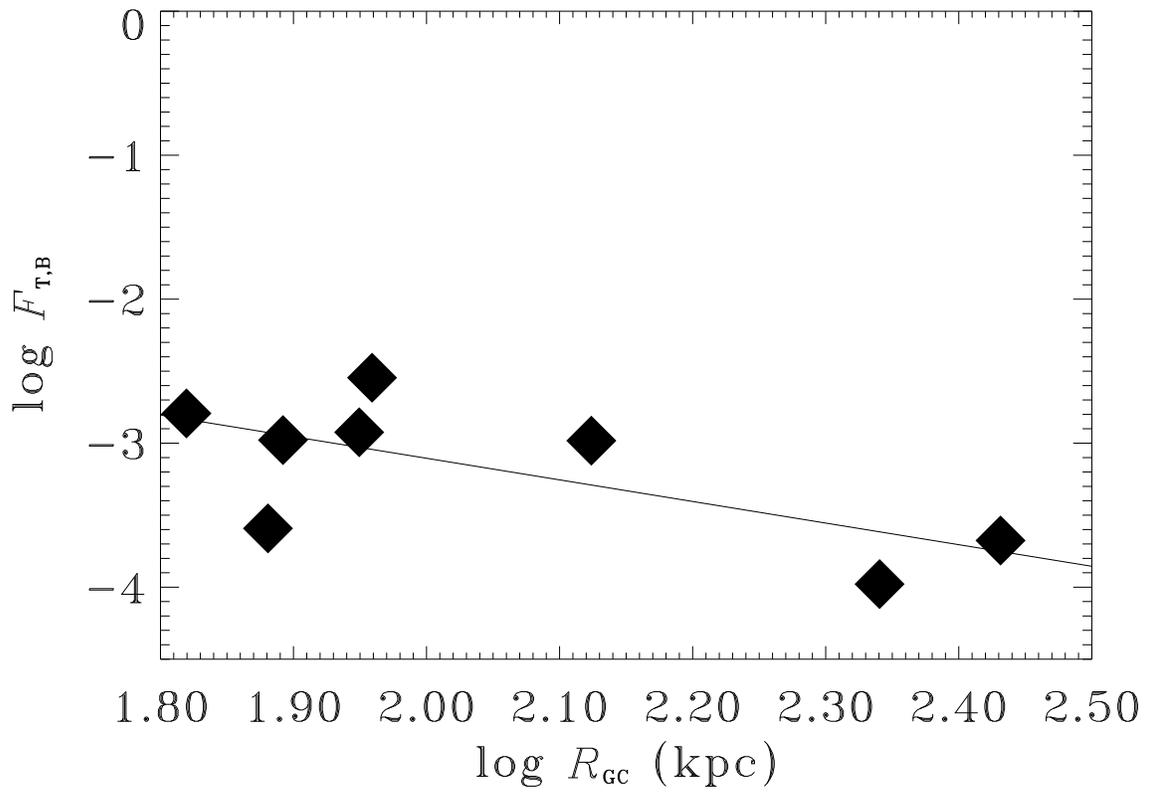
\end{document}